\documentclass[preprint,12pt]{elsarticle}

\usepackage{amssymb}
\usepackage{color}

\journal{Journal of Alloys and Metallurgical Systems}

\begin{document}

\begin{frontmatter}



\title{Metallurgy, superconductivity, and hardness of a new high-entropy alloy superconductor Ti-Hf-Nb-Ta-Re}


\author[inst1]{Takuma Hattori}
\affiliation[inst1]{organization={Department of Electrical Engineering, Faculty of Engineering, Fukuoka Institute of Technology},
            addressline={3-30-1 Wajiro-higashi, Higashi-ku}, 
            city={Fukuoka},
            postcode={811-0295},
            country={Japan}}

\author[inst2]{Yuto Watanabe}

\affiliation[inst2]{organization={Department of Physics, Tokyo Metropolitan University},
            city={Hachioji},
            postcode={192-0397}, 
            country={Japan}}

\author[inst3]{Terukazu Nishizaki}
\affiliation[inst3]{organization={Department of Electrical Engineering, Faculty of Science and Engineering, Kyushu Sangyo University},
            addressline={2-3-1 Matsukadai, Higashi-ku}, 
            city={Fukuoka},
            postcode={813-8503},
            country={Japan}}

\author[inst1]{Koki Hiraoka}

\author[inst1]{Masato Kakihara}

\author[inst2]{Kazuhisa Hoshi}

\author[inst2]{Yoshikazu Mizuguchi}

\author[inst1]{Jiro Kitagawa}

\begin{abstract}
We explored quinary body-centered cubic (bcc) high-entropy alloy (HEA) superconductors with valence electron concentrations (VECs) ranging from 4.6 to 5.0, a domain that has received limited attention in prior research. 
Our search has led to the discovery of new bcc Ti-Hf-Nb-Ta-Re superconducting alloys, which exhibit an interesting phenomenon of phase segregation into two bcc phases with slightly different chemical compositions, as the VEC increases. 
The enthalpy of the formation of each binary compound explains the phase segregation. 
All the alloys investigated were categorized as type-II superconductors, with superconducting critical temperatures ($T_\mathrm{c}$) ranging from 3.25 K to 4.38 K. 
We measured the Vickers microhardness, which positively correlated with the Debye temperature, and compared it with the hardness values of other bcc HEA superconductors. 
Our results indicate that $T_\mathrm{c}$ systematically decreases with an increase in hardness beyond a threshold of approximately 350 HV. 
Additionally, we plotted $T_\mathrm{c}$ vs. VEC for representative quinary bcc HEAs. The plot revealed the asymmetric VEC dependence. 
The correlation between the hardness and $T_\mathrm{c}$, as well as the asymmetric dependence of $T_\mathrm{c}$ on VEC can be attributed to the simultaneous effects of the electronic density of states at the Fermi level and electron-phonon coupling under the uncertainty principle, especially in the higher VEC region.
\end{abstract}




\end{frontmatter}


\section{Introduction}
A high-entropy alloy (HEA) is a type of alloy that consists of multiple elements as its primary constituents, and it differs from conventional alloys that typically have a single principal element with the minor additive elements. 
The high-entropy state of HEAs offers outstanding thermal stability owing to an increased configurational entropy\cite{Biswas:book}. 
HEAs have received significant attention owing to their diverse functionalities, including high fracture toughness, energy storage, thermoelectricity, soft ferromagnetism, high-temperature structural stability, and biocompatibility\cite{Li:PMS2021,Marques:EES2021,Yang:NM2022,Jiang:Science2021,Kitagawa:APLMater2022,Kitagawa:JMMM2022,Xiong:JMST2023,Castro:Metals2021}. 
High-entropy states have been extensively studied in various materials such as oxides, chalcogenides, borides, carbides, and nitrides\cite{Brahlek:APLMater2022,Ying:JACS,Murchie:ACT2022,Wang:AAC2022,Lu:ASS2021}.
The formation of solid solutions in typical HEAs with simple body-centered cubic (bcc) and face-centered cubic (fcc) structures significantly depends on the valence electron concentration (VEC), with VECs ranging from 5.0 to 6.87 for a single bcc phase and exceeding 8.0 for a single fcc phase\cite{Biswas:book}. 
In addition, VECs occasionally affect the physical properties of HEAs. For instance, the Vickers microhardness in bcc and fcc HEAs exhibit a universal relationship with the VECs and form broad peak at the VECs $\sim$ 6.8\cite{Kitagawa:JMMM2022,Tian:IM2015,Kitagawa:JALCOM2022}. 
Furthermore, VEC plays a crucial role in understanding the superconductivity in bcc HEA superconductors, which is a focal point of investigation in the present study.

 Superconductivity is an essential property of HEAs.
The bcc alloy Ta$_{34}$Nb$_{33}$Hf$_{8}$Zr$_{14}$Ti$_{11}$ was the first HEA superconductor reported in 2014\cite{Kozelj:PRL2014}. 
Since then, research on HEA superconductors have become a popular topic\cite{Sun:PRM2019,Kitagawa:Metals2020}.
HEAs exhibit several interesting features. 
The robustness of superconductivity at extremely high pressures has been well established. 
In (TaNb)$_{0.67}$(HfZrTi)$_{0.33}$, the superconducting critical temperature $T_\mathrm{c}$ of 8 K remains almost unchanged even under an extremely high pressure of 180 GPa\cite{Guo:PNAC2017}. 
Additionally, the robustness of the superconductivity has been observed in the presence of magnetic elements \cite{Liu:JALCOM2021}.
Enhancement of the diamagnetic signal in the high-entropy state was confirmed in BiS$_{2}$-based superconductors\cite{Sogabe:SSC2019}.
Recently, Ta-Nb-Hf-Zr-Ti films have been reported to exhibit critical current densities exceeding those of high-field superconducting magnets \cite{Jung:NC2022}. 
Moreover, these films exhibit extremely robust superconductivity under ion irradiation\cite{Jung:NC2022}.  

HEA superconductivity is currently being investigated across various structural types, including bcc\cite{Rohr:PRM2018,Marik:JALCOM2018,Ishizu:RINP,Harayama:JSNM2021,Sarkar:IM2022,Motla:PRB2022,Kitagawa:RHP2022}, hexagonal close-packed (hcp)\cite{Lee:PhysicaC2019,Marik:PRM2019,Browne:JSSC2023}, CsCl-type\cite{Stolze:ChemMater2018}, A15\cite{Wu:SCM2020,Yamashita:JALCOM2021}, NaCl-type\cite{Mizuguchi:JPSJ2019,Yamashita:DalTran2020}, $\alpha$ (or $\beta$)-Mn-type\cite{Stolze:JMCC2018,Xiao:SM2023}, $\sigma$-phase type\cite{Liu:ACS2020}, CuAl$_{2}$-type\cite{Kasen:SST2021}, W$_{5}$Si$_{3}$-type\cite{Liu:PRM2023}, BiS$_{2}$-based, and YBCO-based\cite{Sogabe:SSC2019,Shukunami:PhysicaC2020} structures.
Most HEAs exhibit conventional s-wave BCS-type superconducting properties. 
In bcc Nb-Re-Hf-Zr-Ti and Ta-Nb-Zr-Hf-Ti alloys, the electronic mean free path is considerably smaller than the BCS coherence lengths, categorizing these HEAs within the dirty limit regime, owing to significant atomic disorder\cite{Motla:PRB2022,Leung:SP2022}. 
The extent of the atomic disorder can be quantified using configurational entropy. 
Numerous researchers have explored the influence of configurational entropy on the superconducting properties of HEAs; however, systematic trends have not yet been established. 
Recently, it was proposed that $T_\mathrm{c}$ exhibits a negative correlation with the Debye temperature in several bcc HEAs under a fixed configurational entropy \cite{Kitagawa:JALCOM2022}. 
This behavior can be attributed to the weakened electron-phonon interactions in HEAs with higher Debye temperatures, arising from the uncertainty principle \cite{Kitagawa:JALCOM2022}.
Superconducting Cr$_{5+x}$Mo$_{35-x}$W$_{12}$Re$_{35}$Ru$_{13}$C$_{20}$ has a non-centrosymmetric $\beta$-Mn-type structure\cite{Xiao:SM2023}.
This indicates a significantly enhanced ratio between the upper critical field and Pauli limit in comparison to other $\beta$-Mn-type superconductors\cite{Xiao:SM2023}.
CuAl$_{2}$-type TrZr$_{2}$ (Tr = Fe, Co, Ni, Cu, Rh, and Ir) HEA superconductors exhibit an intriguing phenomenon characterized by anomalous broadening of the specific heat jumps at $T_\mathrm{c}$. This behavior is associated with the microscopic inhomogeneity in the Cooper pair formation\cite{Kasen:SST2021}.
YBCO-type HEAs fall into the category of non-BCS superconductors. 
This structure is characterized by a layered arrangement, and the high-entropy effect at rare-earth sites has been studied in many YBCO HEAs. 
The rare-earth sites are located between the superconducting Cu-O layers. 
An increase in the configurational entropy at rare-earth sites does not significantly affect $T_\mathrm{c}$\cite{Shukunami:PhysicaC2020}.

The bcc HEA superconductors have been extensively studied, with the dependence of $T_\mathrm{c}$ on VEC being typically examined\cite{Marik:JALCOM2018,Ishizu:RINP,Harayama:JSNM2021,Rohr:PNAS2016}.
VEC serves as the reflection of the total density of states of the valence band at the Fermi level, representing a vital factor in determining the $T_\mathrm{c}$ of BCS superconductors. 
A strong correlation between the VEC and $T_\mathrm{c}$ has been observed in binary and ternary superconducting transition-metal alloys, commonly referred to as the Matthias rule\cite{Matthias:PR1955}. 
When plotting $T_\mathrm{c}$ against VEC in binary and ternary superconducting alloys, $T_\mathrm{c}$ exhibits broad peaks at distinct VEC values of approximately 4.6 and 6.6.
Consequently, we utilized VEC as a design parameter to explore new quinary bcc HEAs.
For quinary bcc HEAs, the $T_\mathrm{c}$ of typical superconducting alloys increases as the VEC increases from 4.1 to $\sim$ 4.6–4.7. This behavior is reminiscent of the Matthias rule observed in conventional binary or ternary transition metal alloys.
The $T_\mathrm{c}$ vs. VEC plot for binary or ternary superconducting alloys exhibits a broad peak at VEC of $\sim$ 4.6–4.7. 
Thus, studying bcc HEA superconductors with VEC larger than 4.6 holds great interest, as this region remains relatively unexplored.
The systematic exploration of bcc HEA superconductors with VECs exceeding 4.6 is still in its early stages, with the only bcc HEA superconductors recently reported being Nb-Ta-Mo-Hf-W, Ti-Zr-Nb-Ta-W, and Ti-Zr-Nb-Ta-V superconductors (VEC:4.8 $\sim$ 5.11)\cite{Sobota:PRB2022,Shu:APL2022}.

Transition metal-based quinary bcc HEA superconductors have been reported in various systems, including  Ta-Nb-Hf-Zr-Ti, Nb-V-Hf-Zr-Ti, Ta-V-Hf-Zr-Ti, Nb-Re-Zr-Hf-Ti, Hf-Nb-Ti-V-Zr, Hf-Nb-Ta-Ti-V, Hf-Mo-Nb-Ti-Zr, Nb-Ta-Mo-Hf-W, Ti-Zr-Nb-Ta-W, and Ti-Zr-Nb-Ta-V\cite{Kitagawa:JALCOM2022,Kozelj:PRL2014,Marik:JALCOM2018,Ishizu:RINP,Sarkar:IM2022,Rohr:PNAS2016,Sobota:PRB2022,Shu:APL2022}. 
Thus, the VEC values of these elements appear to be limited to 4, 5, 6, or 7. 
To conveniently achieve an average VEC greater than 4.6, we selected Re (VEC=7) as the constituent element.
The remaining four elements–Ti, Hf, Nb, and Ta–were employed because of their frequent utilization in bcc HEA superconductors, as demonstrated earlier. 
Another crucial design parameter is the $\delta$-parameter, which quantifies the extent of atomic size differences among the constituent elements\cite{Biswas:book,Kitagawa:Metals2020}. 
The $\delta$-parameter can be calculated using the equation,
\begin{equation}
\delta=100\times\sqrt{\sum^{5}_{i=1}c_{i}\left(1-\frac{r_{i}}{\bar{r}}\right)^{2}}
\label{delta}
\end{equation}
Here, $c_{i}$ and $r_{i}$ represent the atomic fraction and radius of the $i$th element, respectively, and $\bar{r}$ denotes the composition-weighted average atomic radius. 
In bcc HEA superconductors, $\delta$ values range from 3.8–10.7\cite{Kitagawa:Metals2020}. 
We confined the atomic fraction of each element between 5 \% and 35 \%, which aligns with one of the definitions of HEA\cite{Biswas:book}.
Within these restriction, we determined the alloy composition such that the VEC ranges from 4.6 to 5.0 and $\delta$ value smaller than 5.0.

In this study, we searched for quinary bcc HEA superconductors with VEC ranging from 4.6 to 5.0 and discovered bcc Ti-Hf-Nb-Ta-Re superconducting alloys. 
These alloys are interesting systems in which phase segregation into two bcc phases occurs with increasing VEC. 
Metallurgical analysis was conducted based on metallographic examination. 
The fundamental superconducting properties were assessed by measuring the electrical resistivity, magnetization, and specific heat. 
The Vickers microhardness, which contains phonon information, was also measured, as the BCS theory suggests that the electron-phonon interaction is one of the decisive factors affecting $T_\mathrm{c}$. 
We discuss the dependence of $T_\mathrm{c}$ on the hardness (or VEC) of several representative bcc HEAs, while considering the effects of the electronic density of states at the Fermi level and electron-phonon coupling under the uncertainty principle.

\section{Materials and Methods}
Polycrystalline samples were prepared using the constituent elements, Ti (99.9 \%), Hf (98 \%), Nb (99.9 \%), Ta (99.9 \%), and Re (99.99 \%), in a homemade arc furnace under an Ar atmosphere, as shown in Table \ref{tab:table1}. 
Each sample underwent multiple flips and remelting to ensure homogeneity, and the samples were finally quenched on a water-chilled Cu hearth. 
The samples received neither heat treatment nor deformation such as rolling.
The VEC was calculated using the following equation:
\begin{equation}
\mathrm{VEC}=\sum^{5}_{i=1}c_{i}\mathrm{VEC}_{i}
\label{vec}
\end{equation}
Here, VEC$_{i}$ is the VEC value associated with the $i$th element. 
The VEC$_{i}$ values were four for Ti and Hf, five for Nb and Ta, and seven for Re. 
The $\delta$-parameter values were as follows: 4.46 for VEC=4.6, 4.24 for VEC=4.7, 3.91 for VEC=4.8, 3.42 for VEC=4.9, and 2.7 for VEC=5.0. 
For the calculation, atomic radii data were sourced from the literature \cite{Miracle:AM2017}.

Room-temperature X-ray diffraction (XRD) patterns were obtained using a Shimadzu XRD-7000L X-ray diffractometer with Cu-K$\alpha$ radiation. 
Scanning electron microscopy (SEM) images were obtained using a JEOL JSM-7100F field-emission scanning electron microscope (FE-SEM). 
An energy-dispersive X-ray (EDX) spectrometer equipped with FE-SEM was employed to evaluate the chemical composition in each sample area. 
The EDX spectrometer was also used to obtain elemental mappings.

A quantum-design MPMS3 SQUID magnetometer was used to measure the temperature dependence of the dc magnetization $M$($T$) and isothermal magnetization curve. 
The electrical resistivity $\rho$ was measured using a four-probe method with a quantum-design PPMS apparatus. 
The specific heat was also measured using PPMS equipment. 
The Vickers microhardness was measured using a Shimadzu HMV-2T microhardness tester under an applied load of 2.94 N, with a 10 s holding time under a diamond indenter.

\section{Results and Discussion}
\subsection{Structural and Metallographic Characterizations}
\begin{figure}
\begin{center}
\includegraphics[width=1\linewidth]{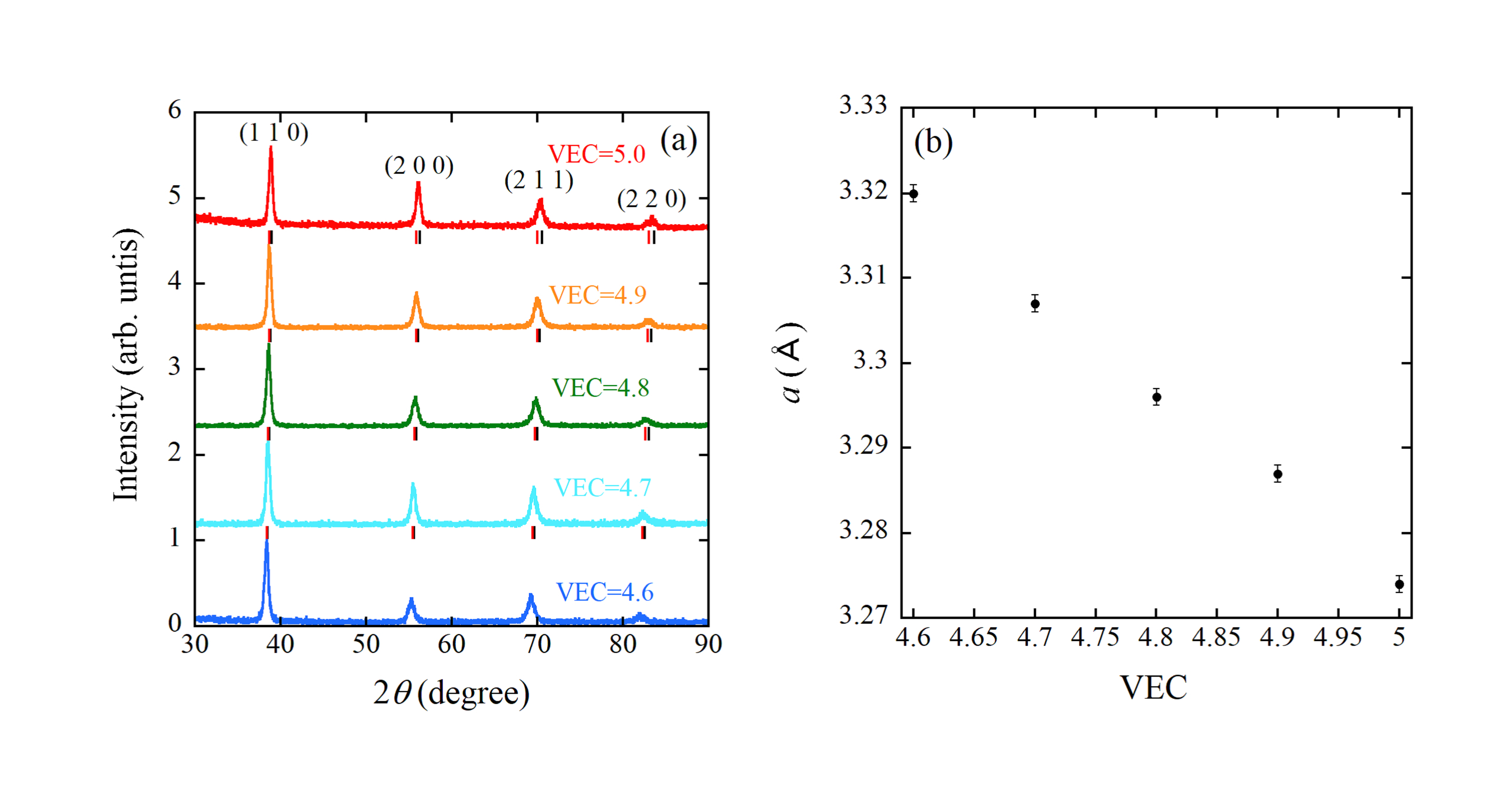}
\caption{\label{fig1}(a) XRD patterns and (b) VEC dependence of the lattice parameters of Ti-Hf-Nb-Ta-Re alloys. In (a), the black and red bars indicate the simulated XRD positions for segregated phases with VECs higher and lower than the nominal VEC, respectively (Table  \ref{tab:table1}).}
\end{center}
\end{figure}

Figure \ref{fig1}(a) shows the XRD patterns of the prepared alloys, which can be readily indexed to the bcc structure with no conspicuous impurity phases, as indicated by the Miller indices in the figure. 
The lattice parameters obtained by the least-squares method are listed in Table \ref{tab:table1} and plotted against VEC in Fig.\hspace{1mm}\ref{fig1}(b). 
The systematic decrease in the lattice parameter with increasing VEC can be attributed to differences in the atomic radii of the constituent elements. 
As VEC increases from 4.6 to 5.0, the concentration of Ti+Hf decreased instead of an increase in Ta or Nb atomic fraction. 
The atomic radii of Ti, Hf, Nb, and Ta are 1.4615,  1.5775, 1.429, and 1.43 \AA, respectively\cite{Miracle:AM2017}.
Hence, the substitution of Ti+Hf with Nb or Ta results in a reduction in the lattice parameter.

\begin{table}
\caption{\label{tab:table1}%
Starting atomic composition, nominal VEC, lattice 
 parameter $a$, and chemical composition determined by the EDX measurement of each prepared sample.}
 \begin{center}
 {\scriptsize
\begin{tabular}{cccc}
\hline
starting composition & VEC & {\it a} (\AA) & determined composition \\
\hline
(Ti$_{35}$Hf$_{25}$)(Nb$_{25}$Ta$_{5}$)Re$_{10}$ & 4.6 & 3.320(1) & (Ti$_{33.2(9)}$Hf$_{25.4(4)}$)(Nb$_{24.9(8)}$Ta$_{6.5(4)}$)Re$_{10.0(1)}$: VEC 4.61 \\
(Ti$_{30}$Hf$_{20}$)(Nb$_{35}$Ta$_{5}$)Re$_{10}$ & 4.7 & 3.307(1) & (Ti$_{27(1)}$Hf$_{18.3(5)}$)(Nb$_{38(1)}$Ta$_{7.5(5)}$)Re$_{10.1(8)}$: VEC 4.75 \\
 &  &  & (Ti$_{30.6(4)}$Hf$_{22.8(5)}$)(Nb$_{30.6(7)}$Ta$_{5.7(1)}$)Re$_{10.3(2)}$: VEC 4.67 \\
(Ti$_{25}$Hf$_{15}$)(Nb$_{35}$Ta$_{15}$)Re$_{10}$ & 4.8 & 3.296(1) & (Ti$_{22(1)}$Hf$_{12.7(6)}$)(Nb$_{37.2(6)}$Ta$_{17.1(5)}$)Re$_{11.1(5)}$: VEC 4.88 \\
 &  &  & (Ti$_{26(1)}$Hf$_{18(1)}$)(Nb$_{33(1)}$Ta$_{12(1)}$)Re$_{10.4(3)}$: VEC 4.77 \\
(Ti$_{20}$Hf$_{10}$)(Nb$_{35}$Ta$_{25}$)Re$_{10}$ & 4.9 & 3.287(1) & (Ti$_{16(1)}$Hf$_{8.2(9)}$)(Nb$_{36.7(9)}$Ta$_{28(1)}$)Re$_{10.7(5)}$: VEC 4.97 \\
 &  &  & (Ti$_{22(1)}$Hf$_{12(1)}$)(Nb$_{35(1)}$Ta$_{21(1)}$)Re$_{9.9(4)}$: VEC 4.86 \\
(Ti$_{15}$Hf$_{5}$)(Nb$_{35}$Ta$_{35}$)Re$_{10}$ & 5.0 & 3.274(1) & (Ti$_{12(1)}$Hf$_{4.0(6)}$)(Nb$_{36.4(5)}$Ta$_{36(1)}$)Re$_{11.5(4)}$: VEC 5.07 \\
 &  &  & (Ti$_{21.5(3)}$Hf$_{9.4(3)}$)(Nb$_{36.3(7)}$Ta$_{23.1(3)}$)Re$_{9.6(5)}$: VEC 4.88 \\
\hline
\end{tabular}
}
\end{center}
\end{table}

\begin{figure}
\begin{center}
\includegraphics[width=1.2\linewidth]{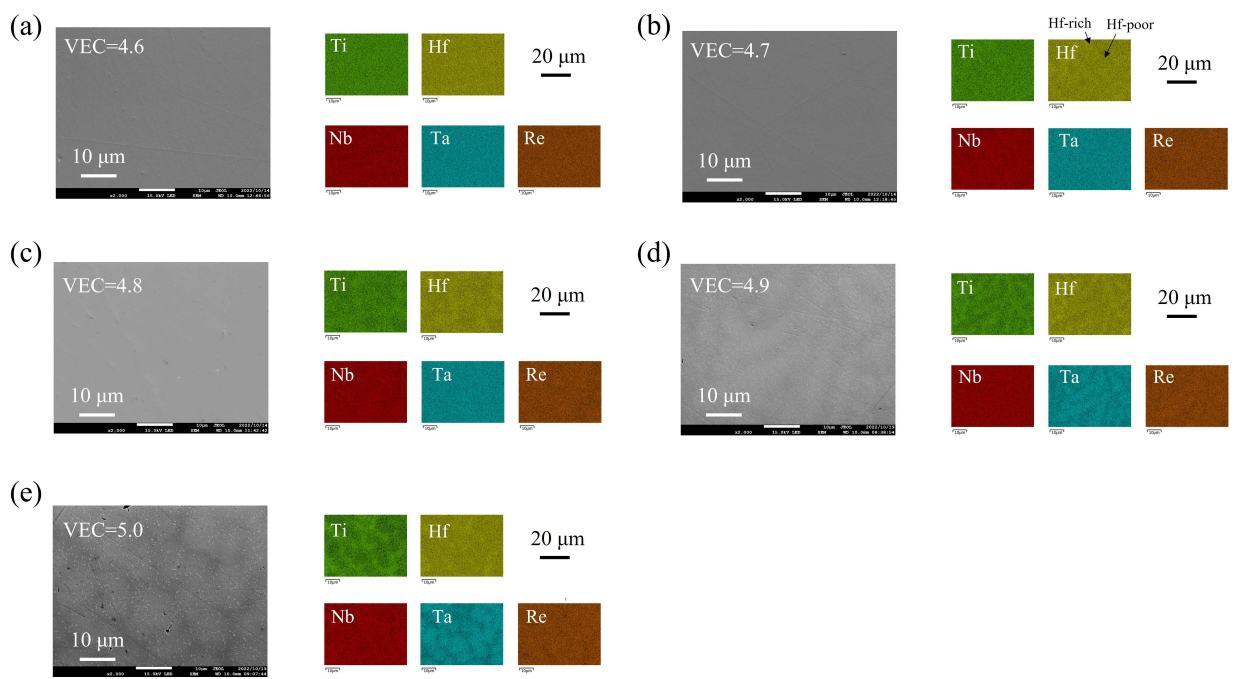}
\caption{\label{fig2} SEM images and elemental mappings of the samples with VECs of (a) 4.6, (b) 4.7, (c) 4.8, (d) 4.9, and (e) 5.0.}
\end{center}
\end{figure}

Figures \ref{fig2}(a)–(e) show the SEM images and corresponding elemental maps for all alloys. 
The SEM image of the sample with a VEC of 4.6 exhibits no noticeable secondary phase, and each elemental mapping demonstrates a homogeneous distribution, as shown in Fig.\hspace{1mm}\ref{fig2}(a). 
The chemical composition determined using EDX agreed with the initial composition (Table \ref{tab:table1}). 
As discussed later, the sharp superconductive transition in the specific heat strongly supports the single-phase nature of the sample with a VEC of 4.6.
Increasing the VEC induces phase segregation, which is evident in samples with VEC = 4.9 and 5.0 (Figs.\hspace{1mm}\ref{fig2}(d) and (e)).
In each sample, both the SEM image and elemental mapping revealed phase segregation in each sample. The bright phase in the SEM image of the sample with a VEC of 4.9 is rich in Nb and Ta and poor in Ti and Hf, whereas the dark phase exhibits the reverse tendency. 
As shown in Table \ref{tab:table1}, the VEC value calculated using the chemical composition of the bright (dark) phase is higher (lower) than the nominal value of 4.9. 
Similar results were obtained for the sample with a VEC of 5.0. 
Although the samples with VEC = 4.7 and 4.8 exhibited no contrasted SEM images, weak inhomogeneous elemental distributions were detected in Ti and Hf (Figs.\hspace{1mm}\ref{fig2}(b) and (c)). 
We have verified the chemical compositions of the samples with VECs of 4.7 and 4.8, based on the elemental mappings (e.g., Hf-rich and Hf-poor regions noted in Fig.\hspace{1mm}\ref{fig2}(b)). 
In each sample, the results of EDX analyses suggested the segregation into Nb- and Ta-rich phases and Ti- and Hf-rich phases. 
The weak manifestation of phase segregation in the sample with a VEC of 4.7 or 4.8 is consistent with the observation of a slightly rounded specific heat jump, as discussed below.

The XRD patterns of the phase-segregated samples was composed of two phases with slightly varying VEC values. 
In these samples, the chemical compositions of both phases were comparable, and as a result, they constituted a bcc structure. 
The average value of the lattice parameters for the samples with VEC = 4.7, 4.8, 4.9, or 5.0 is shown in Fig.\hspace{1mm}\ref{fig1}(b). 
Therefore, based on a linear approximation of the data plot shown in Fig.\hspace{1mm}\ref{fig1}(b), the lattice parameters of the two phases for each phase-segregated sample can be estimated by utilizing the actual VEC values determined through EDX. 
The XRD positions of the two phases, one with a VEC higher and the other with a VEC lower than the nominal VEC, are shown in Fig.\hspace{1mm}\ref{fig1}(a) by the black and red bars, respectively. 
This indicates that the XRD patterns of the phase-segregated samples can be explained by the superposition of the two phases.

Next, we examine the mechanism underlying the phase segregation phenomenon based on the enthalpy of formation ($\Delta$$H_\mathrm{f}$). 
We utilized $\Delta$$H_\mathrm{f}$ data\cite{Troparevsky:PRX2015} for each binary compound with an equimolar ratio of the constituent elements, as presented in Table \ref{tab:table2}. 
Our analysis revealed that Ti and Nb exhibited attractive interactions with Hf and Ta, respectively, whereas Ti and Hf atoms exhibited repulsive interactions with Nb and Ta atoms. 
These results explain the segregation observed in the (Ti, Hf)- and (Nb, Ta)-rich phases. 
Furthermore, Re atoms tend to form alloys with all the other elements because of their significantly negative $\Delta$$H_\mathrm{f}$ values. 
Consequently, the inhomogeneity of the Re distribution is not as pronounced. 
To elucidate the phenomenon of phase segregation in more detail, we performed calculations to determine the enthalpy of mixing ($\Delta H{_\mathrm{mix}}$) of the quinary alloy using the relationship in \cite{Biswas:book}
\begin{equation}
\Delta H_{\mathrm{mix}}=4\sum^{5}_{i=1, i\neq j}\Delta H_\mathrm{f(i,j)} c_{i} c_{j} 
\label{hmix}
\end{equation}
Here, $\Delta H_\mathrm{f(i,j)}$ denotes the enthalpy of the binary phase between the $i$th and $j$th elements, as listed in Table \ref{tab:table2}. The resulting $\Delta H{_\mathrm{mix}}$ values for the samples with VECs of 4.6, 4.7, 4.8, 4.9, and 5.0 were found to be -163, -153, -142, -138, and -142 (meV/atom), respectively. 
These findings indicate the diminishing stability of the equiatomic solid solution with increasing VEC. 
As previously mentioned, $\Delta H_\mathrm{f(Ti,Hf)}$ and $\Delta H_\mathrm{f(Nb,Ta)}$ exhibit negative values, while $\Delta H_\mathrm{f(Ti,Nb)}$, $\Delta H_\mathrm{f(Ti,Ta)}$, $\Delta H_\mathrm{f(Hf,Nb)}$, and $\Delta H_\mathrm{f(Hf,Ta)}$ exhibit positive values. Consequently, within the context of an unstable equiatomic solid solution, segregation into (Ti, Hf)-rich and (Nb, Ta)-rich phases becomes energetically favorable. This segregation aligns with the observed phase segregation.

\begin{table}
\caption{\label{tab:table2}%
Enthalpies of each binary compound. The unit is meV/atom. Data was obtained from \cite{Troparevsky:PRX2015}.}
\begin{center}
\begin{tabular}{c|ccccc}
\hline
 & Ti & Hf & Nb & Ta & Re \\
\hline
Ti & 0 & -10 & 11 & 31 & -189 \\
Hf & -10 & 0 & 23 & 49 & -407 \\
Nb & 11 & 23 & 0 & -10 & -202 \\
Ta & 31 & 49 & -10 & 0 & -226 \\
Re & -189 & -407 & -202 & -226 & 0 \\
\hline
\end{tabular}
\end{center}
\end{table}

\subsection{Superconducting Properties}
\begin{figure}
\begin{center}
\includegraphics[width=1\linewidth]{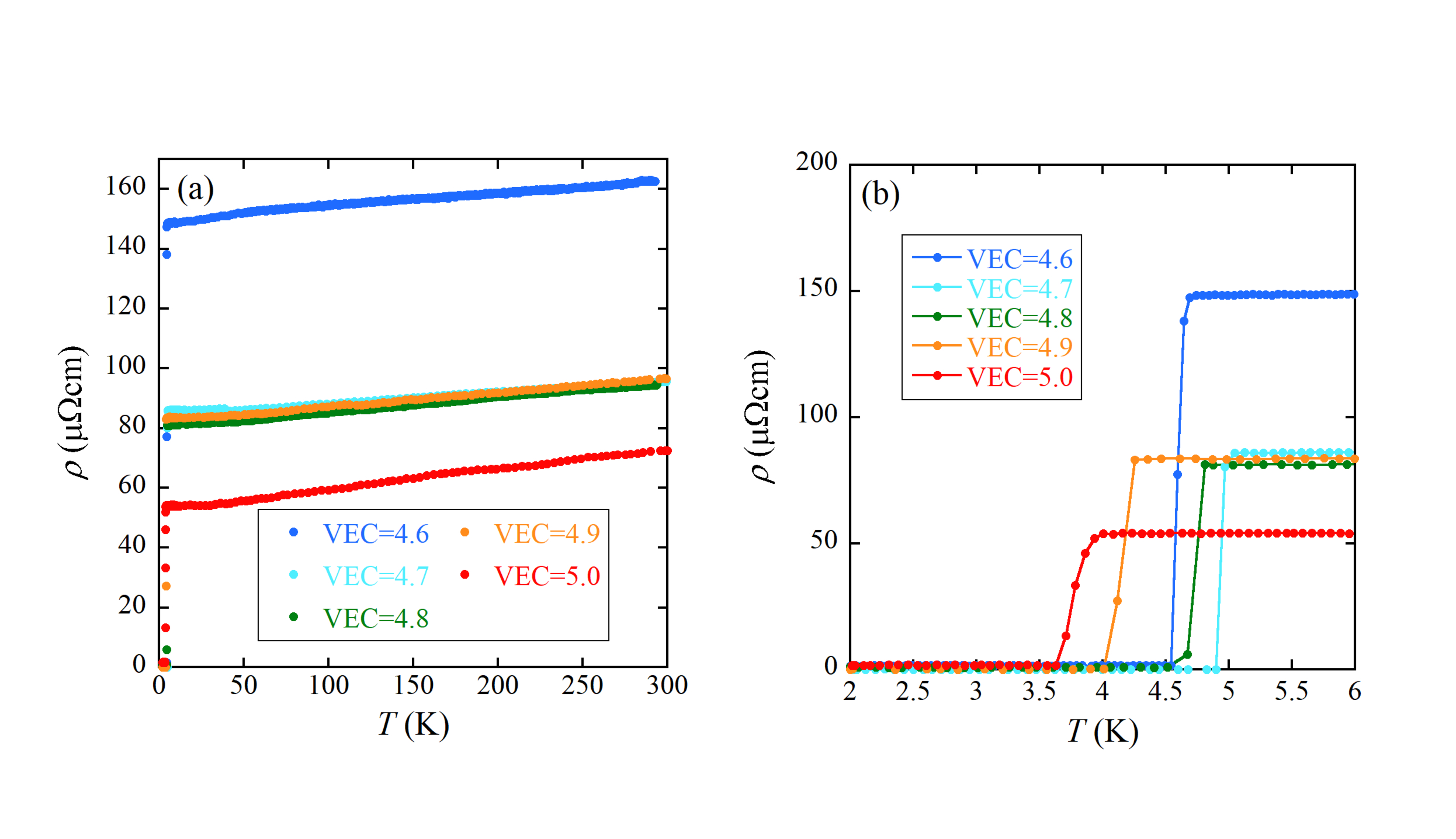}
\caption{\label{fig3} (a) Temperature dependences of the electrical resistivities of the Ti-Hf-Nb-Ta-Re alloys. (b) Low temperature electrical resistivity of the Ti-Hf-Nb-Ta-Re alloys.}
\end{center}
\end{figure}

Figure \ref{fig3}(a) illustrates the temperature dependence of $\rho$ for all the alloys. 
The order of magnitude of $\rho$ for each sample indicated good metallicity, whereas atomic disorder was responsible for the weak temperature dependence. 
Figure \ref{fig3}(b) shows the low-temperature $\rho$ behaviors, which exhibit sharp drops to zero-resistivity states at $T_\mathrm{c}$. 
The resistivity drop of each alloy starts slightly above $T_\mathrm{c}$, which is determined by the magnetization or specific heat, as reported for several HEA superconductors\cite{Kitagawa:JALCOM2022,Sarkar:IM2022,Stolze:ChemMater2018}.

\begin{figure}
\begin{center}
\includegraphics[width=0.7\linewidth]{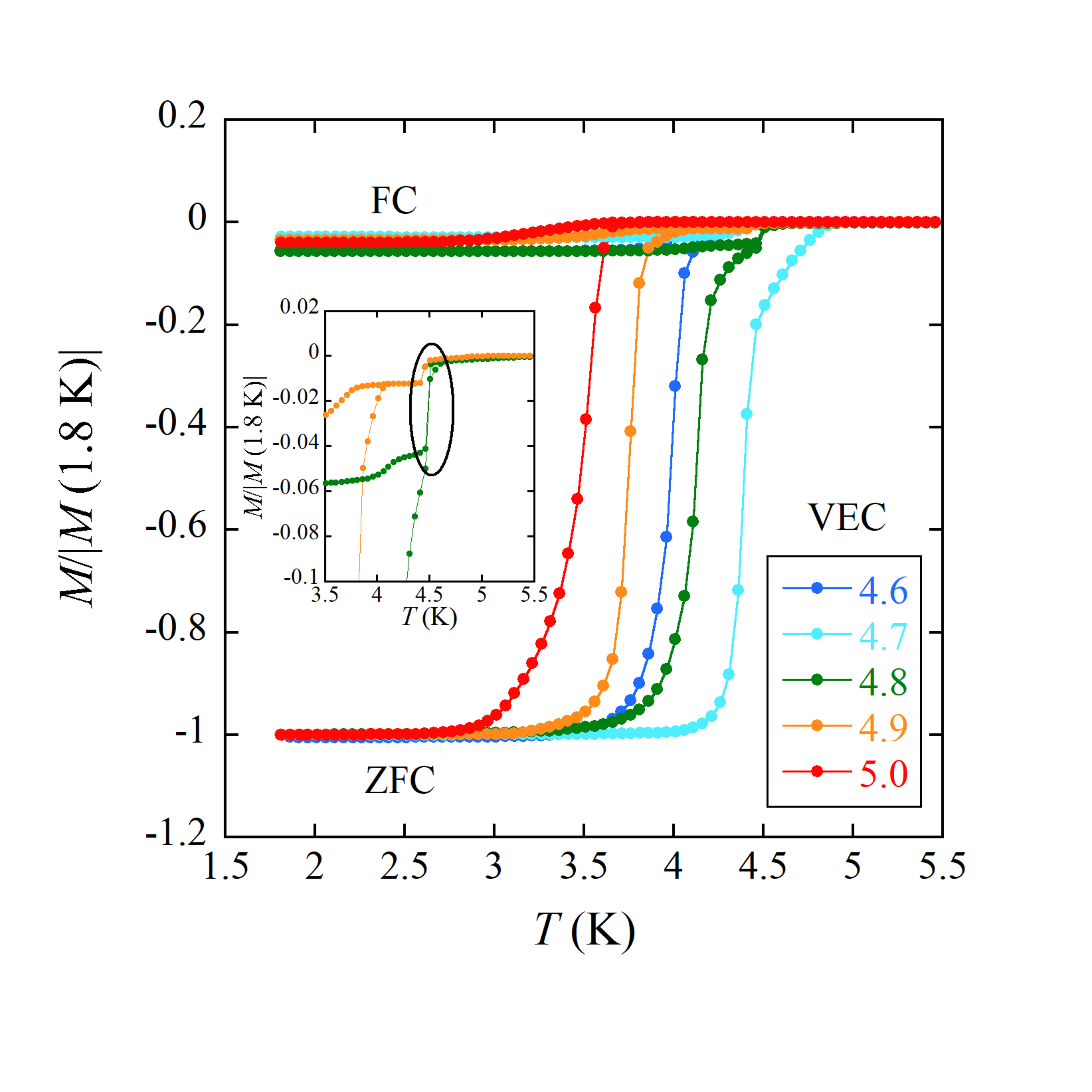}
\caption{\label{fig4}Temperature dependences of the $M$ values of Ti-Hf-Nb-Ta-Re alloys. $M$ is normalized by the absolute value of $M$ at 1.8 K. The inset is the expanded view of $M$ around 4.5 K for the samples with  VECs of 4.8 and 4.9.}
\end{center}
\end{figure}

The results of temperature-dependent $M$ measured under zero-field cooled (ZFC) and field cooled (FC) conditions, under an external field $\mu_{0}H$ of 0.4 mT, are presented in Fig.\hspace{1mm}\ref{fig4}. 
Each ZFC data point shows a diamagnetic signal due to superconductivity, whereas the FC data point indicates flux pinning in the sample. 
The samples with VECs of 4.8 and 4.9 possess additional small sharp drops in $M$ at approximately 4.5 K (the solid ellipsoid in the inset of Fig.\hspace{1mm}\ref{fig4}). 
These drops are due to the parasitic superconducting phase resulting from phase segregation. 
This parasitic phase differed from the two segregated phases in each sample, and the specific heat with no anomaly at 4.5 K indicates a negligible amount of the parasitic phase.

\begin{figure}
\begin{center}
\includegraphics[width=1\linewidth]{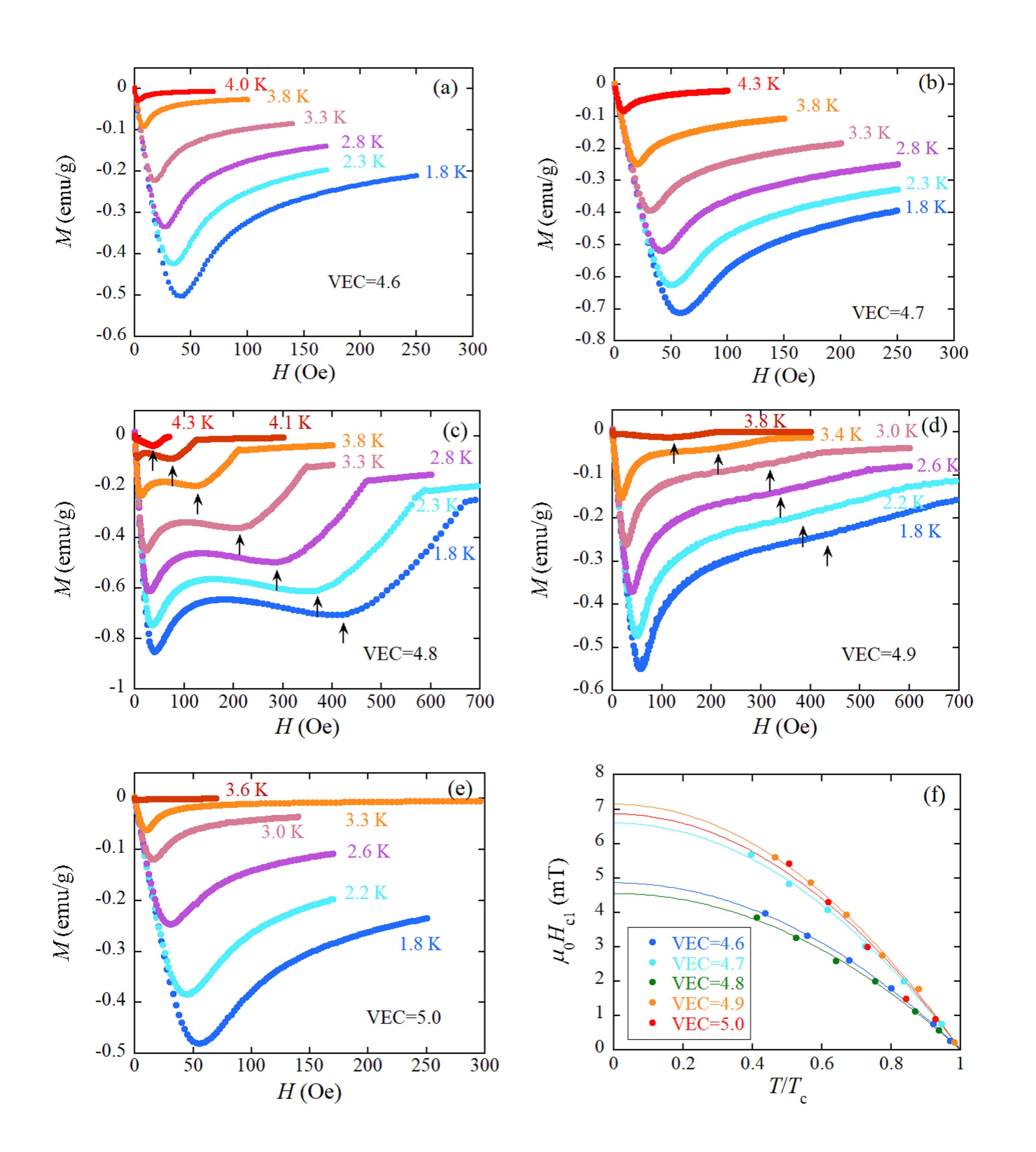}
\caption{\label{fig5}Field-dependent magnetization of the samples with VECs of (a) 4.6, (b) 4.7, (c) 4.8, (d) 4.9, and (e) 5.0. (f) Temperature dependences of the lower critical field of the Ti-Hf-Nb-Ta-Re alloys. The solid curves show the fitting results using eq. (\ref{eq:hc1}).}
\end{center}
\end{figure}

To evaluate the lower critical field $H_\mathrm{c1}$, the $M$–$H$ curves at lower $H$ were measured at different temperatures, as summarized in Figs.\hspace{1mm}\ref{fig5}(a)-(e). 
$H_\mathrm{c1}$ is defined as the $H$ value where a negative peak is observed. 
The samples with VECs of 4.8 and 4.9 show broad anomalies at $H$ higher than $H_\mathrm{c1}$, denoted by arrows. 
These anomalies correlated with the appearance of a parasitic superconducting phase.
The sample with a VEC of 4.8 and relatively larger $M$ drop at 4.5 K in the inset of Fig.\hspace{1mm}\ref{fig4}, shows more obvious anomalies in the $M$-$H$ curves (see Fig.\hspace{1mm}\ref{fig5}(c)). 
The temperature dependence of $H_\mathrm{c1}$ can be reproduced using the Ginzburg-Landau equation given as follows:
\begin{equation}
H_\mathrm{c1}(T)=H_\mathrm{c1}(0)\left(1-t^{2}\right),
\label{eq:hc1}
\end{equation}
where $t$ is the reduced temperature $\frac{T}{T_\mathrm{c}}$ (Fig.\hspace{1mm}\ref{fig5}(f)).
The $\mu_{0}H_\mathrm{c1}$(0) values are listed in Table \ref{tab:table3}.

\begin{figure}
\begin{center}
\includegraphics[width=1\linewidth]{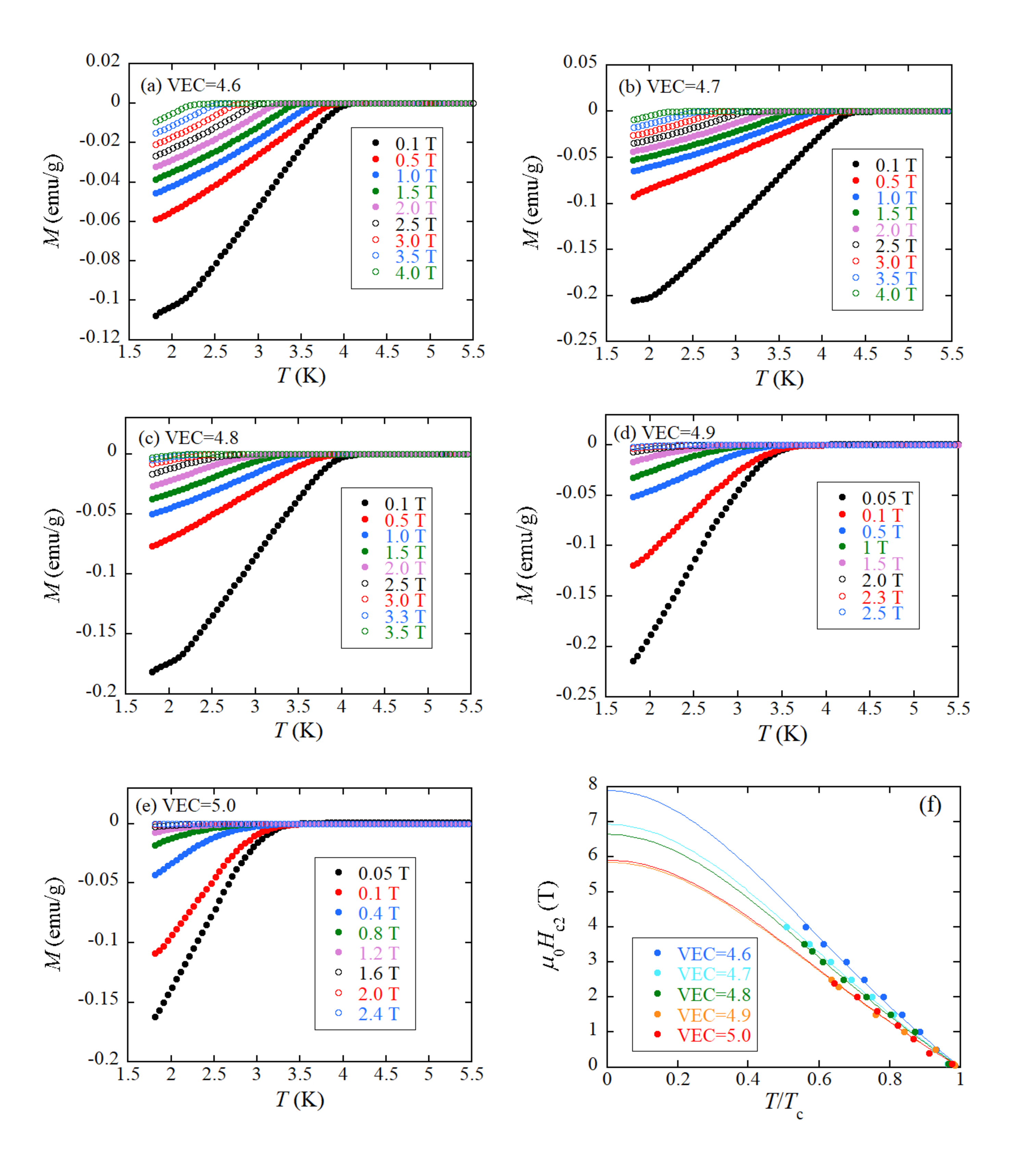}
\caption{\label{fig6}Temperature-dependent ZFC magnetization under external fields denoted in figure for the samples with VECs of (a) 4.6, (b) 4.7, (c) 4.8, (d) 4.9, and (e) 5.0. (f) Temperature dependences of the upper critical fields of the Ti-Hf-Nb-Ta-Re alloys. The solid curves show the fitting results using eq. (\ref{eq:hc2}).}
\end{center}
\end{figure}

We also measured the temperature dependence of ZFC $M$ for various $\mu_{0}H$ values to estimate the upper critical field $H_\mathrm{c2}$, as shown in Figs.\hspace{1mm}\ref{fig6}(a)-(e). 
The onset of the diamagnetic signal is defined as $T_\mathrm{c}$, and $\mu_{0}H_\mathrm{c2}$ for each sample is plotted as a function of $T$ in Fig.\hspace{1mm}\ref{fig6}(f). 
The temperature dependence of $\mu_{0}H_\mathrm{c2}$ can be explained using the following formula:
\begin{equation}
H_\mathrm{c2}(T)=H_\mathrm{c2}(0)\left(\frac{1-t^{2}}{1+t^{2}}\right)
\label{eq:hc2}
\end{equation}
The obtained $\mu_{0}H_\mathrm{c2}$ values are summarized in Table \ref{tab:table3}.

For each alloy, the Ginzburg-Landau coherence length $\xi_\mathrm{GL}$(0) is calculated as, $\xi_\mathrm{GL}(0)=\sqrt{\frac{\Phi_{0}}{2\pi\mu_{0}H_\mathrm{c2}(0)}}$, where $\Phi_{0}$ denotes the magnetic flux quantum of 2.07$\times$10$^{-15}$ Wb. 
The resulting values are 6.5, 6.9, 7.0, 7.5, and 7.5 nm for the VEC values of 4.6, 4.7, 4.8, 4.9, and 5.0, respectively. 
The magnetic penetration depth $\lambda_\mathrm{GL}$(0) can be obtained by using the relation, $\mu_{0}H_\mathrm{c1}(0)=\frac{\Phi_{0}}{4\pi\lambda_\mathrm{GL}(0)^{2}}\mathrm{ln}\frac{\lambda_\mathrm{GL}(0)}{\xi_\mathrm{GL}(0)}$. 
The $\lambda_\mathrm{GL}$(0) values extracted from the data are 370, 308, 380, 291, and 297 nm for the VEC values of 4.6, 4.7, 4.8, 4.9, and 5.0, respectively. 
The Ginzburg-Landau parameter $\kappa_\mathrm{GL}$=$\lambda_{GL}(0)/\xi_{GL}(0)$ was then calculated. The resulting Ginzburg-Landau parameter values were 57, 45, 54, 39, and 40 for the VEC values of 4.6, 4.7, 4.8, 4.9, and 5.0, respectively. 
All $\kappa_\mathrm{GL}$ values were found to be greater than 1/$\sqrt{2}$, indicating that all the alloys are type-II superconductors.

\begin{figure}
\begin{center}
\includegraphics[width=1\linewidth]{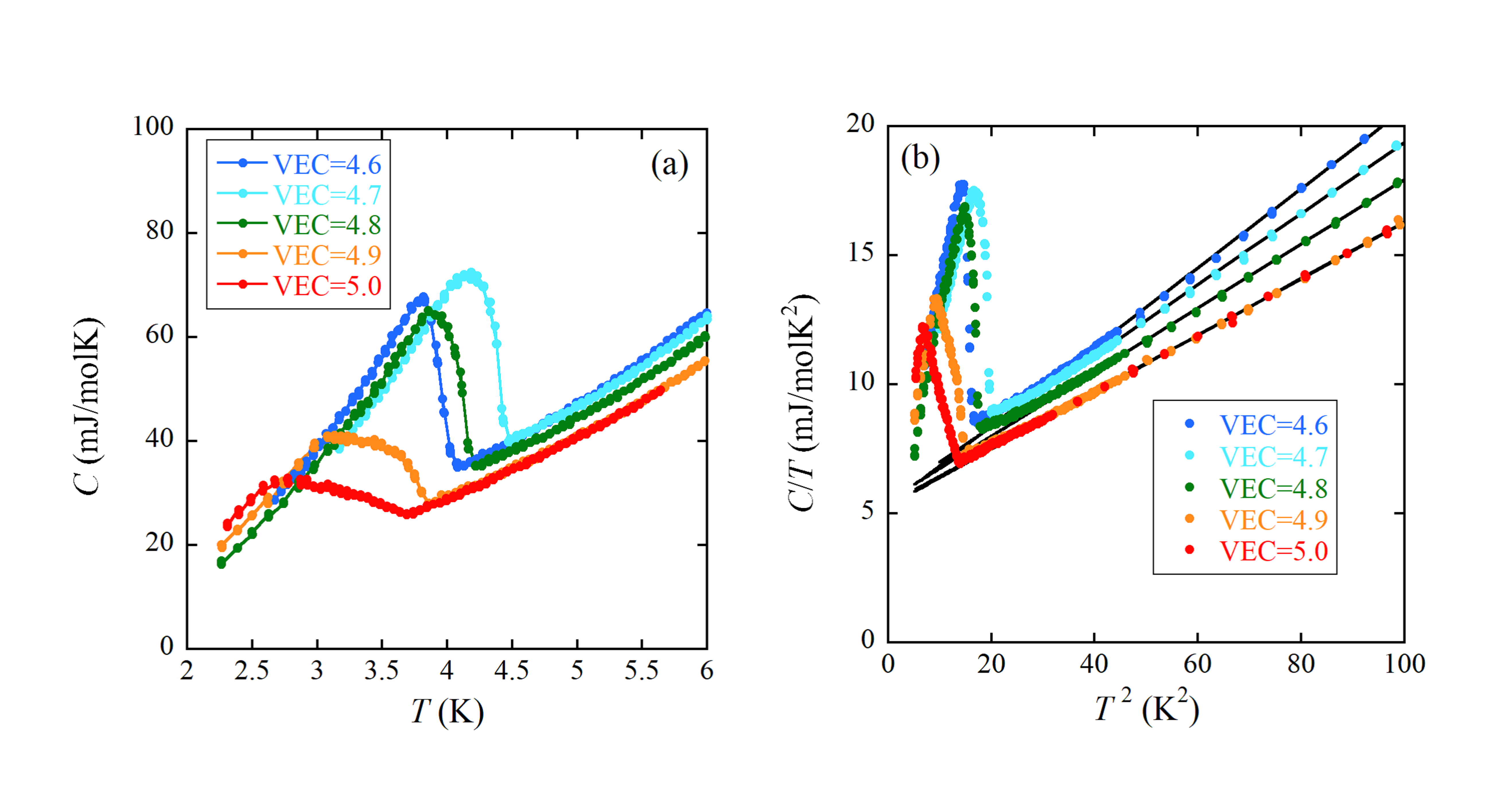}
\caption{\label{fig7}(a) Temperature dependences of the $C$ values of the Ti-Hf-Nb-Ta-Re alloys. (b) $C/T$ vs. $T^{2}$ plots of the Ti-Hf-Nb-Ta-Re alloys. Each solid line shows the fitting result using eq. (\ref{eq:cp}).}
\end{center}
\end{figure}

Figure \ref{fig7}(a) shows the comparison of the temperature dependences of the $C$ ($C$($T$)) values for all alloys. 
Each sample exhibited the bulk nature of superconductive transition, and $T_\mathrm{c}$ was defined as the midpoint of the transition at each $C$($T$) (Table \ref{tab:table3}). 
As VEC increases from 4.6 to 5.0, the superconductive transition gradually broadens owing to the progressive growth of the phase segregation. 
In each phase-segregated sample, the chemical compositions of the two phases with a bcc structure were similar, and each phase underwent a superconductive transition.
In Fig.\hspace{1mm}\ref{fig7}(b), $C/T$ is plotted as a function of $T^{2}$, and each alloy data above $T_\mathrm{c}$ can be fitted by the expression,
\begin{equation}
\frac{C}{T}=\gamma+\beta T^{2},
\label{eq:cp}
\end{equation}
where $\gamma$ and $\beta$ are the Sommerfeld coefficient and lattice contribution, respectively. 
The Debye temperature $\theta_\mathrm{D}$, was derived from the $\beta$ value using the equation, $\theta_\mathrm{D}=\left(\frac{12\pi^{4}RN}{5\beta}\right)^{1/3}$, where the number of atoms per formula unit $N$ = 1, and $R$ is the gas constant.
The values of $\gamma$ and $\theta_\mathrm{D}$ for each alloy are presented in Table \ref{tab:table3}. 
As the VEC values increases from 4.6 to 4.7, $\gamma$ increases, and a further increase in VEC after a threshold of 4.7 tends to decrease $\gamma$. 
However, $\theta_\mathrm{D}$ systematically increases with increasing VEC.

\begin{table}
\caption{\label{tab:table3}%
Superconducting parameters and Vickers microhardness of Ti-Hf-Nb-Ta-Re. The nominal compositions of alloys are (Ti$_{35}$Hf$_{25}$)(Nb$_{25}$Ta$_{5}$)Re$_{10}$ for VEC=4.6, (Ti$_{30}$Hf$_{20}$)(Nb$_{35}$Ta$_{5}$)Re$_{10}$ for VEC=4.7, (Ti$_{25}$Hf$_{15}$)(Nb$_{35}$Ta$_{15}$)Re$_{10}$ for VEC=4.8, (Ti$_{20}$Hf$_{10}$)(Nb$_{35}$Ta$_{25}$)Re$_{10}$ for VEC=4.9, and (Ti$_{15}$Hf$_{5}$)(Nb$_{35}$Ta$_{35}$)Re$_{10}$ for VEC=5.0, respectively. $T_\mathrm{c}$ is determined by specific heat measurement.}
\begin{center}
\begin{tabular}{cccccc}
\hline
VEC & 4.6 & 4.7 & 4.8 & 4.9 & 5.0 \\
\hline
$T_\mathrm{c}$ (K) & 3.95 & 4.38 & 4.10 & 3.62 & 3.25 \\
$\gamma$ (mJ/mol$\cdot$K$^{2}$) & 5.33 & 5.61 & 5.52 & 5.30 & 5.37 \\
$\theta_\mathrm{D}$ (K) & 233 & 241 & 250 & 260 & 261 \\
$\mu_{0}H_\mathrm{c1}$ (mT) & 4.87 & 6.60 & 4.56 & 7.15 & 6.87 \\
$\mu_{0}H_\mathrm{c2}$ (T) & 7.90 & 6.94 & 6.65 & 5.85 & 5.91 \\
Hardness (HV) & 438.5(7.0) & 427.6(4.5) & 445.6(8.5) & 460.0(9.5) & 466.2(6.0) \\
\hline
\end{tabular}
\end{center}
\end{table}

The intriguing aspect lies in the broad superconducting transition of $C$($T$) observed in the samples with VECs of 4.9 and 5.0, characterized by prominent phase segregation. 
Certain bcc HEA superconductors exhibit phase segregation, and their $C$($T$) data have been previously reported\cite{Sarkar:IM2022,Krnel:materials2022}. 
For instance, the equiatomic ScHfNbTaTi compound decomposes into bcc superconducting and hcp non-superconducting phases\cite{Krnel:materials2022}.
The superconducting phase displayed a distinct and sharp $C$($T$) transition at $T_\mathrm{c}$=6.1 K, whereas the hcp phase had no discernible effect on the sharpness of the specific heat jump at $T_\mathrm{c}$. 
Another example is the equiatomic HfNbTaTiV compound (VEC=4.6), which exhibits a dendritic structure\cite{Sarkar:IM2022}. 
Both the dendritic phase (VEC=4.57) and inter-dendritic phase (VEC=4.48) manifested as bcc superconductors. 
Although the difference between the VECs the two phases is similar to that of the present alloy (VEC=4.9 or 5.0), the $C$($T$) value of HfNbTaTiV shows a sharp transition at $T_\mathrm{c}$=4.37 K. 
In comparison to ScHfNbTaTi and HfNbTaTiV, the broad superconducting transition of the $C$($T$) value observed in the current Ti-Hf-Nb-Ta-Re alloy with a VEC of 4.9 or 5.0 appears to be a rare phenomenon.
This raises intriguing questions regarding the relationship between the microstructure and superconducting properties of HEAs.

\subsection{Hardness}
\begin{figure}
\begin{center}
\includegraphics[width=1.0\linewidth]{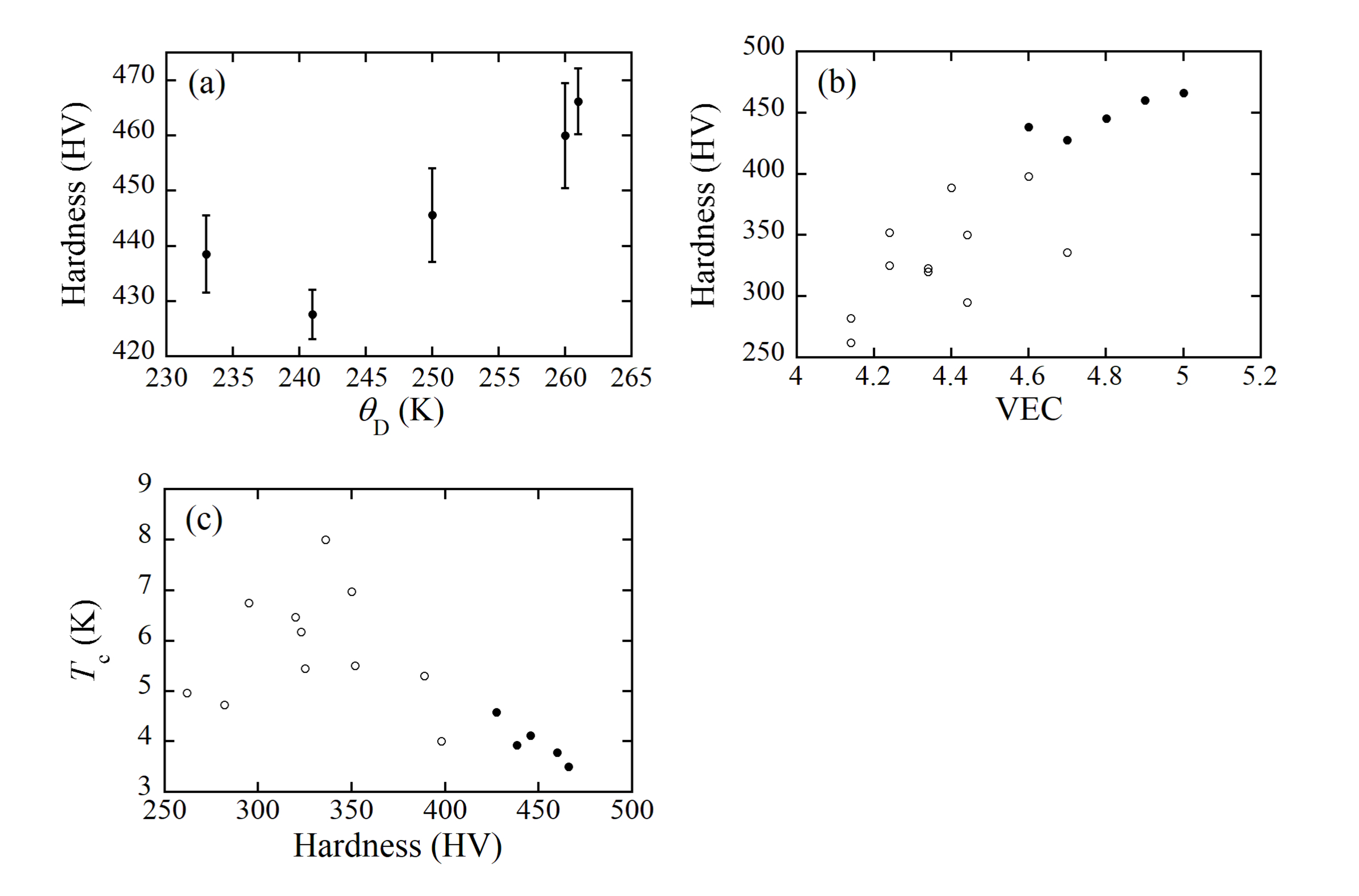}
\caption{\label{fig8}(a) Vickers microhardness vs. $\theta_\mathrm{D}$ plot for the Ti-Hf-Nb-Ta-Re alloys. (b) VEC dependence of the Vickers microhardness of the bcc HEA superconductors. Data drawn by open circles from \cite{Kitagawa:JALCOM2022}. (c) Hardness dependence of the $T_\mathrm{c}$ values of the bcc HEA superconductors used for Vickers microhardness examinations. Data represented by open circles were obtained from \cite{Kitagawa:JALCOM2022}.}
\end{center}
\end{figure}

Table \ref{tab:table3} lists the Vickers microhardness values of the examined alloys. 
Figure \ref{fig8}(a) depicts the relationship between the Vickers microhardness and $\theta_\mathrm{D}$, where $\theta_\mathrm{D}$ represents the highest frequency of the lattice vibration, which reflects the interatomic bonding strength. 
A material with a higher $\theta_\mathrm{D}$ exhibits enhanced bonding forces between atoms, which results in increased hardness, as observed in Fig.\hspace{1mm}\ref{fig8}(a). 
Recent first-principles calculations conducted on transition metal binary or high-entropy alloys, which possess elemental combinations akin to Ti-Hf-Nb-Ta-Re, further support the positive correlation between hardness and $\theta_\mathrm{D}$\cite{Jiang:FED2021,Ru:MTC2022}.
Considering that $\theta_\mathrm{D}$ generally reflects the hardness of a material, it is reasonable to anticipate that an increase in $\theta_\mathrm{D}$ corresponds to an increase in the Vickers microhardness.
Figure \ref{fig8}(a) reveals a positive correlation between hardness and $\theta_\mathrm{D}$, suggesting that hardness can serve as an alternative to $\theta_\mathrm{D}$ in explaining the electron-phonon interactions. 
Several reports have demonstrated that the Vickers microhardness of numerous bcc and fcc HEAs depends on the VEC values \cite{Kitagawa:JMMM2022,Tian:IM2015,Kitagawa:JALCOM2022}.
In bcc HEAs with VECs of $\sim$ 4–5.5, an increase in VEC tends to result in a higher hardness because the hard elements are present at VEC values of $\sim$ 5–7. 
This behavior is reinforced in Fig.\hspace{1mm}\ref{fig8}(b), where the hardness is plotted as a function of VEC, along with our prior findings\cite{Kitagawa:JALCOM2022} on typical bcc HEA superconductors.

Figure \ref{fig8}(c) shows the plot of $T_\mathrm{c}$ vs. hardness for the alloys utilized in Fig.\hspace{1mm}\ref{fig8}(b), to examine the impact of hardness on HEA superconductivity.
Investigations into the hardness characteristics of bcc HEA superconductors have not been undertaken by any other research group, to the best of our knowledge.
At hardness values greater than approximately 350 HV, $T_\mathrm{c}$ decreases as the hardness increases, owing to two factors. 
According to the BCS theory, $T_\mathrm{c}$ depends on the electronic density of states at the Fermi level and electron-phonon interaction. 
Thus, the first factor is the decrease in the electronic density of states at the Fermi level with increasing hardness, because an HEA with a hardness exceeding 350 HV is situated on a VEC larger than approximately 4.5, as depicted in Fig.\hspace{1mm}\ref{fig8}(b). 
The Matthias rule for conventional binary or ternary transition metal alloys proposes that $T_\mathrm{c}$ forms a broad peak at a VEC of 4.6.
The Matthias rule, which is widely recognized, emphasizes the significant influence of the electronic density of states at the Fermi level in determining $T_\mathrm{c}$. This rule states that the electronic density of states at the Fermi level is systematically depressed as VEC increases beyond approximately 4.6. 
Therefore, an HEA with hardness exceeding 350 HV may also have a reduced electronic density of states, leading to a lower $T_\mathrm{c}$. 
Ti-Hf-Nb-Ta-Re exhibits a positive correlation between $T_\mathrm{c}$ and $\gamma$, where $\gamma$ reflects the magnitude of the electronic density of states at the Fermi level, thus supporting the first factor. 
The second factor is the decline in $T_\mathrm{c}$ owing to the weakened electron-phonon interactions through the uncertainty principle. 
A broad phonon band is typically expected in the disordered atomic state \cite{Kormann:npjCM2017,Mizuguchi:MTP2023}. 
An HEA with higher hardness possesses higher $\theta_\mathrm{D}$, which leads to a broader phonon band. 
This results in a shorter phonon lifetime according to the uncertainty principle $\Delta E\Delta t\geq\hbar/2$, where $\Delta E$ is the energy uncertainty associated with band broadening, $\Delta t$ is the lifetime, and $\hbar$ is Planck’s constant. 
Therefore, the electron-phonon coupling of an HEA with higher hardness is weakened\cite{Kitagawa:JALCOM2022}, leading to the lower $T_\mathrm{c}$.

\subsection{$T_\mathrm{c}$ vs. VEC plot}
\begin{figure}
\begin{center}
\includegraphics[width=0.6\linewidth]{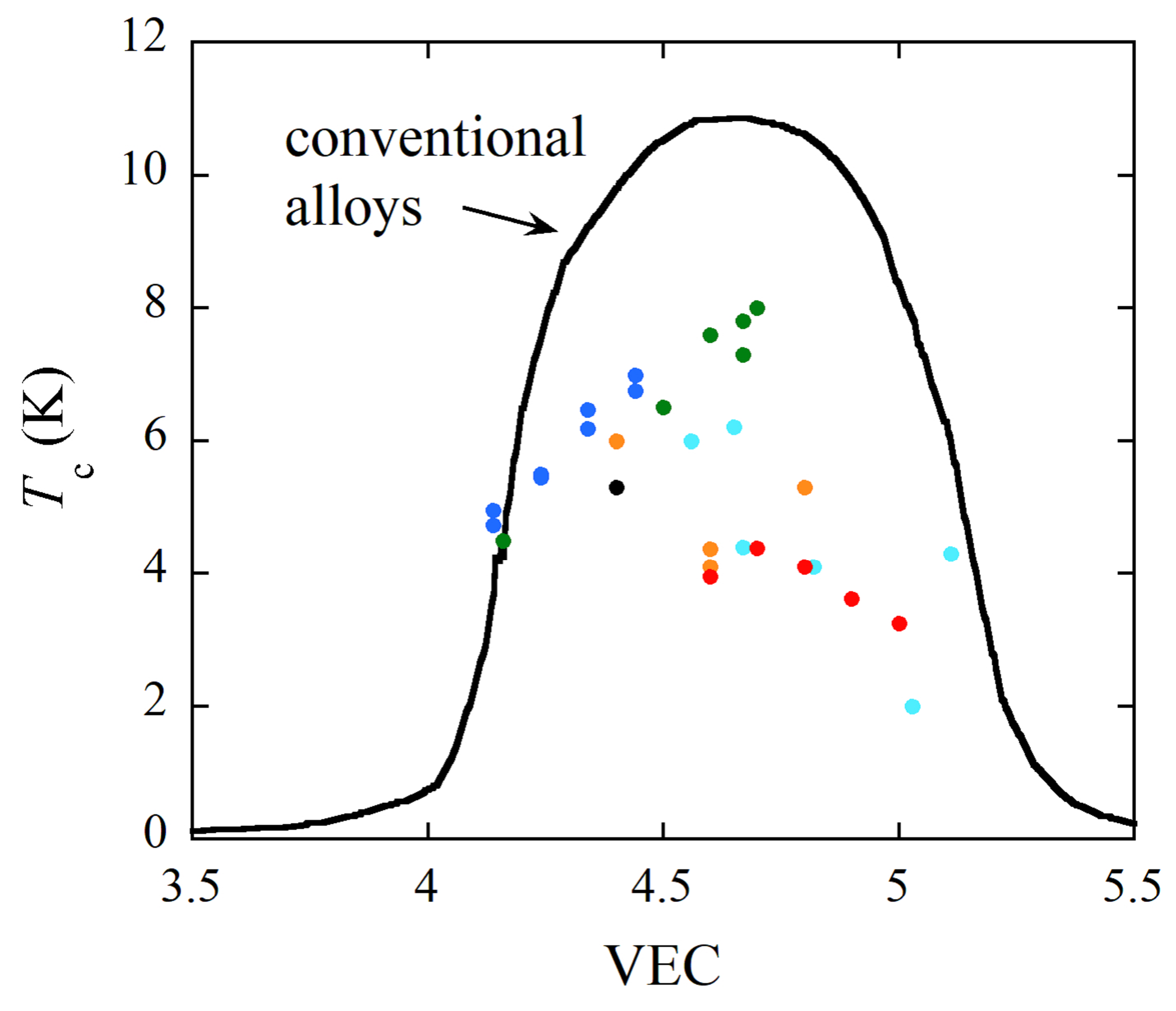}
\caption{\label{fig9}VEC dependence of the $T_\mathrm{c}$ values of typical quinary bcc HEA superconductors. The correspondence between color and HEA is as follows: green: non-equimolar Hf-Nb-Ta-Ti-Zr; blue: Al-Nb-Ti-V-Zr; black: Hf$_{21}$Nb$_{25}$Ti$_{15}$V$_{15}$Zr$_{24}$; orange: HfNbTaTiZr, HfNbReTiZr, HfNbTaTiV, and HfMoNbTiZr; light blue: Nb-Ta-Mo-Hf-W, Ti-Zr-Nb-Ta-W, and Ti-Zr-Nb-Ta-V; and red: Ti-Hf-Nb-Ta-Re. The solid line represents the Matthias rule of conventional binary or ternary transition metal alloys.}
\end{center}
\end{figure}

Here, we present a comparison of the superconducting properties of the quinary bcc HEA superconductors, with VECs ranging from 4.1 to 5.3. 
The $T_\mathrm{c}$ data are plotted against VEC in Fig.\hspace{1mm}\ref{fig9}.
The green, blue, and black symbols represent non-equimolar Hf-Nb-Ta-Ti-Zr, Al-Nb-Ti-V-Zr, and Hf$_{21}$Nb$_{25}$Ti$_{15}$V$_{15}$Zr$_{24}$, respectively\cite{Ishizu:RINP,Harayama:JSNM2021,Rohr:PNAS2016}. 
The orange symbols denote the equiatomic quinary bcc HEA superconductors\cite{Kitagawa:JALCOM2022,Marik:JALCOM2018,Sarkar:IM2022,Vrtnik:JALCOM2017} of HfNbTaTiZr, HfNbReTiZr, HfNbTa-TiV, and HfMoNbTiZr. 
Studies corresponding the VECs of 5.0 or more are indicated by light blue (Nb-Ta-Mo-Hf-W, Ti-Zr-Nb-Ta-W, and Ti-Zr-Nb-Ta-V)\cite{Sobota:PRB2022,Shu:APL2022} and red (this work) symbols. 
The solid curve in the plot represents the Matthias rule for conventional binary or ternary transition-metal alloys. 
According to this rule, the electronic density of states at the Fermi level, which is closely related to VEC, is the decisive factor for $T_\mathrm{c}$.
While the solid curve is symmetric with respect to VEC, the quinary bcc HEA superconductors show an asymmetric trend.
This contrasted behavior means that a factor other than the electronic density of states at the Fermi level plays a role in the $T_\mathrm{c}$ determination of HEAs, especially at VECs larger than $\sim$4.6.
Considering that $T_\mathrm{c}$ depends on the electronic density of states at the Fermi level and electron-phonon interaction, the weight of the electron-phonon interaction in determining $T_\mathrm{c}$ increases in HEAs with VECs exceeding 4.6. 
As mentioned in the previous subsection, the uncertainty principle in HEAs may contribute to a weakened electron-phonon coupling strength at VECs exceeding 4.6, leading to an additional reduction in $T_\mathrm{c}$. 
Therefore, we speculate that the simultaneous effect of the reduced electronic density of states at the Fermi level and weakened electron-phonon coupling through the uncertainty principle in the higher VEC region gives rise to the hardness dependence of $T_\mathrm{c}$ and asymmetric $T_\mathrm{c}$ vs. VEC plot.

We discuss the correlation between the Vickers microhardness and $\theta_\mathrm{D}$ based on the $T_\mathrm{c}$ vs. VEC plot. 
The correlations between Vickers microhardness and $\theta_\mathrm{D}$, presented in Fig.\hspace{1mm}\ref{fig8}(a), suggests that a harder sample exhibits a higher $\theta_\mathrm{D}$. 
Hardness is linked to VEC, and within the current alloy system, hardness increases with increasing VEC, as shown in Fig.\hspace{1mm}\ref{fig8}(b). 
The plot of $T_\mathrm{c}$ vs. VEC in Fig.\hspace{1mm}\ref{fig9} illustrates that the $T_\mathrm{c}$ of Ti-Hf-Nb-Ta-Re decreases with increasing VEC. 
Consequently, harder samples tend to exhibit a lower $T_\mathrm{c}$, indicating that a higher $\theta_\mathrm{D}$ results in a lower $T_\mathrm{c}$. This negative correlation between $\theta_\mathrm{D}$ and $T_\mathrm{c}$ contrasts with the positive correlation typically observed in conventional BCS superconductors. 
A previous study\cite{Kitagawa:JALCOM2022} that compared quinary bcc HEA superconductors also revealed a negative correlation between $\theta_\mathrm{D}$ and $T_\mathrm{c}$, suggesting that this could be a common characteristic of bcc HEA superconductors.

\section{Summary}
We have discovered the superconducting alloy of bcc Ti-Hf-Nb-Ta-Re with VEC values ranging from 4.6 to 5.0 are all type-II superconductors with $T_\mathrm{c}$= $\sim$3.25–4.38 K. 
A significant feature of this system is the emergence of phase segregation when the VEC surpasses 4.7, which can be explained by the enthalpy of the formation of each binary compound. 
Because of this phase segregation, $C$($T$) data exhibited a broad superconducting transition in the samples with VECs of 4.9 or 5.0. 
We confirmed a positive correlation between the Vickers microhardness and $\theta_\mathrm{D}$. 
Furthermore, we examined the hardness dependence of $T_\mathrm{c}$ for various quinary bcc HEAs and discovered a systematic reduction in $T_\mathrm{c}$, when the hardness exceeded approximately 350 HV. 
The plot of $T_\mathrm{c}$ vs. VEC for the representative quinary bcc HEAs reveals an asymmetric VEC dependence. 
Contrary to the Matthias rule for conventional binary or ternary transition metal alloys, the $T_\mathrm{c}$ values of HEAs are significantly suppressed at high VEC values.
The reduction in the electronic density of states at the Fermi level and weakened electron-phonon coupling through the uncertainty principle could be responsible for the hardness dependence of $T_\mathrm{c}$ at hardness values exceeding 350 HV and substantial suppression of $T_\mathrm{c}$ in the high VEC region.


\section*{Acknowledgments}
T.N. acknowledges the support from a Grant-in-Aid for Scientific Research (KAKENHI) (Grant No. 20K03867) and the Advanced Instruments Center of Kyushu Sangyo University. Y.M. acknowledges the support from a Grant-in-Aid for Scientific Research (KAKENHI) (Grant No. 21H00151). J.K. acknowledges the support from a Grant-in-Aid for Scientific Research (KAKENHI) (Grant No. 23K04570) and the Comprehensive Research Organization of the Fukuoka Institute of Technology. 


\end{document}